\documentstyle[eqsecnum,multicol,epsfig,aps,prl,array]{revtex}
\newcommand{\bleq}{\ifpreprintsty
		   \else
		   \end{multicols}\widetext \vspace*{-3.5ex}{\tiny
		   
		\noindent\begin{tabular}[t]{c|}
		   \parbox{0.493\hsize}{~} \\ \hline \end{tabular}}
				      \fi}
\newcommand{\eleq}{\ifpreprintsty
		   \else
		   {\tiny\hspace*{\fill}\begin{tabular}[t]{|c}\hline
		    \parbox{0.49\hsize}{~} \\
		    \end{tabular}}\vspace*{-2.5ex}\begin{multicols}{2}
		    \narrowtext
		    \fi}
\newcommand{\bcols}{\ifpreprintsty\else\begin{multicols}{2} 
	\narrowtext\fi}
\newcommand{\ecols}{\ifpreprintsty\else\end{multicols}\fi}

\draft
\widetext
\begin{document}
\title{Structure of densified amorphous germanium dioxide} 
\author{Matthieu Micoulaut}
\address{Laboratoire de Physique Th{\'e}orique des Liquides,
Universit{\'e} Pierre et Marie Curie, Boite 121\\
4, Place Jussieu, 75252 Paris Cedex 05, France\\}
\date{\today}
\maketitle
\begin{abstract}
\par
Classical molecular dynamics simulations are used to study the
structure of densified germanium dioxide ($GeO_2$). It is found that
the coordination number of germanium changes with increasing density
(pressure) while pressure released systems
 do not exhibit a marked change in local structure as
compared to the virgin system. The structural modification with
pressure appears to be stepwise and gradually affects long range 
(through the reduction of the long-range correlations as seen from the
shift of the first sharp diffraction peak), intermediate range (by
angular reduction) and finally short range structure (by tetrahedron 
distorsion).\par
{Pacs:} 61.43.Fs-46.30.Cn
\end{abstract}
\par
\bcols
Structural transitions in minerals are known
to take place under various geological conditions\cite{r1,r2}. In the Earth's
interior, silicate and alumino-silicate melts change their local
structure causing strong density and viscosity modifications in magmas\cite{r3}
and silicon
exhibits at high pressure a change of its coordination
number\cite{r4}. 
These results
have been mostly obtained with high pressure-temperature
experiments\cite{r5} but also with computer simulations\cite{r6} 
reproducing extreme
conditions. Both have inferred the nature of
these structural transitions, the structure and the
phase portrait of the liquid
state. However, while a majority of studies have been devoted to the
silica and silicate
chemistry, little has been done to elucidate the corresponding
behavior in germanium dioxide ($GeO_2$) even though this material is a 
structural analog of silica in many respects:
both materials exhibit at ordinary conditions a tetrahedral local
structure, they can also exist in $\alpha$ as well as $\beta$ quartz
phases\cite{r7} and
the change of germanium coordination from four to six also 
occurs at high pressure\cite{r8}. The structure of {\em in situ}
densified or permanently densified $GeO_2$ remains however
controversial. While the global increase with
pressure of the 
germanium-oxygen distance in $GeO_2$ has been
related from x-ray diffraction \cite{r8} with the conversion of
tetrahedral Ge(4) into octahedral 
Ge(6), it seems that this structural change is reversible as no Ge(6)
is found in decompressed samples\cite{Hannon}, a situation which does
not occur\cite{r8} in the decompressed rutilelike $c-GeO_2$ 
The interest in
germanium coordination change has been also motivated by the
observation of the so-called ``{\em germanate
anomaly}'' which corresponds to a maximum in density and refractive
index when $15-16~mol\%$ $Na_2O$ are added into the basic $GeO_2$
network former\cite{anomaly}. These binary systems
have been investigated by various spectroscopic tools
\cite{r11}-\cite{r12} and it is suggested that the increasing presence
of $GeO_6$ octahedra (Ge(6)) within the network is
responsable of the anomaly. On
the other hand, micro-Raman\cite{r13} applied on the same
systems suggests that the anomaly is due to a massive
conversion of 4-membered rings into more close-packed ones such as
3-membered rings, with no, or at least very few, Ge(6) present.
\par 
A preliminary task if one wishes to describe density induced structural
changes in germanates, is first
to understand how the basic network former $GeO_2$ changes with
densification. While crystalline phases of $GeO_2$ have been studied
from numerical calculations\cite{r14,r15}, we are not aware of any
published result on simulated liquid and amorphous $GeO_2$. It is
therefore of striking interest to see what Molecular Dynamics (MD) can
tell about a certain number of experimental open questions which
remain at this stage. How does
densification affect the local structure in the amorphous material ? 
Which thermodynamical 
quantity (temperature, Pressure)
controls mostly the existence of a 6-fold coordinated germanium ? Does
octahedral germanium exists in pressure released (permanently
densified) $GeO_2$ ? This
Letter attempts to address some of these basic issues by providing the first
MD study of amorphous germania. The results show several main
features. They demonstrate the reversible nature of pressure
induced changes and no major difference with the virgin structure is
found upon pressure release, while the local coordination number of the
germanium increases smoothly from ~4 to 5.5 with applied pressure, a result
that would be accessible from {\em in situ} experiments as already
realized for silica\cite{Hemley92}. Furthermore, several r\'egimes of
densification can be clearly identified: a first r\'egime [$P\leq 1.8~GPa$]
during which no global change in the local structure is found to occur
whereas the increase in the position of the first sharp diffraction
peak (FSDP) suggests a global reduction of the longer range
correlations. In the window $1.8~GPa\leq P\leq 2.8 GPa$, a sharp decrease of
the intertetrahedral bond angle permits to densify the structure with no
distorsion of the basic $GeO_{4/2}$ tetrahedron. This leads
to the buckling of the network connected tetrahedra.  Finally, for larger
pressures, distorsion of the tetrahadron sets in. As a
consequence, additional constraints appear for $P>3GPa$ which produce
a global stiffening of the network\cite{Thorpe,Dove}
\par
The system consists of 256 germanium and 512 oxygen atoms interacting
via a Born-Huggins-Mayer type potential which has been fitted in the
case of $GeO_2$ by Oeffner and Elliot to recover the crystalline
phases of $GeO_2$ and its vibrational spectra\cite{r14}. 
The atoms have been
first confined in a cubic box of length $L=23.044 \AA$ in order to recover
the experimental value of the density ($\rho_g=3.66~g.cm^{-3}$). After having
thermalized the system at $3000~K$ for 20000 time steps (20 ps) the
system has been cooled to $300~K$ with a linear cooling schedule at a
quench rate of $2.5\times10^{12}~K.s^{-1}$. Integration has been done using
a leap-frog Verlet algorithm. Various configurations
(positions and velocities) have been saved at different temperatures
which have been used as starting configurations for production runs of
$10^5$ steps.
\begin{figure}
\begin{center}
\epsfig{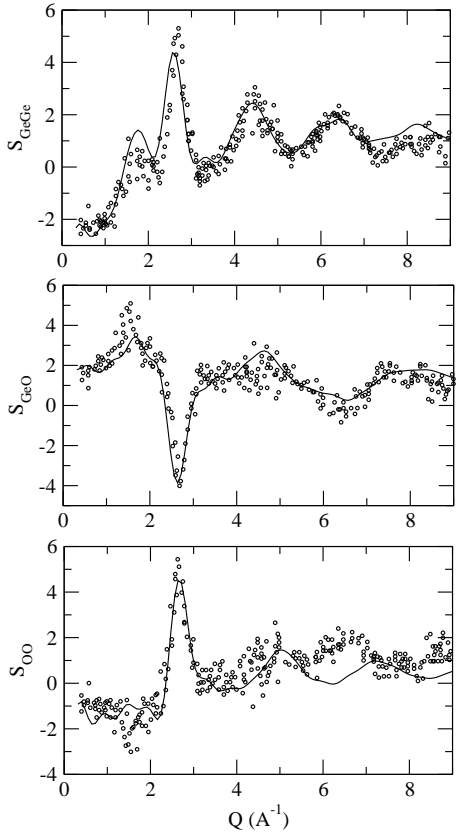}
\end{center}
\caption{Partial structure factors $S_{ij}(Q)$ of vitreous germania at
$300~K$ from MD simulations (solid line) at $\rho=\rho_g$ 
compared to experimental results (circles) from [11].}
\end{figure}
Densification has been realized reproducing experimental
conditions, i.e. starting from an initial configuration at $T=300~K$
and $\rho_g$ and increasing the density during $10^4$ time steps. At the 
density $\rho_g$, the glass
transition temperature was about $1660~K$, and as expected, the $T_g$
shifts to the higher temperatures with increasing density ($1894~K$ at
$\rho=1.1\rho_g$). The sample data of Price and
co-workers\cite{r12} has been almost recovered, i.e. after MD pressure 
release from 15.16GPa determined from the virial\cite{detail}. The
final density of the decompressed system at
zero pressure was
$\rho=4.5~g.cm^{-3}=1.25\rho_g$. The corresponding configuration (termed in
the following as ``{\em permanently densified system}'') was taken
for comparison with the experimental study\cite{Price2003} of
permanently densified $GeO_2$
\par
The results for the partial structure factors $S_{ij}(Q)$ are displayed in
Fig. 1 at $\rho_g=3.66~g.cm^{-3}$ which show a fair agreement with the
experimental results obtained from a combination of x-ray
data\cite{r12}.
\begin{figure}
\begin{center}
\epsfig{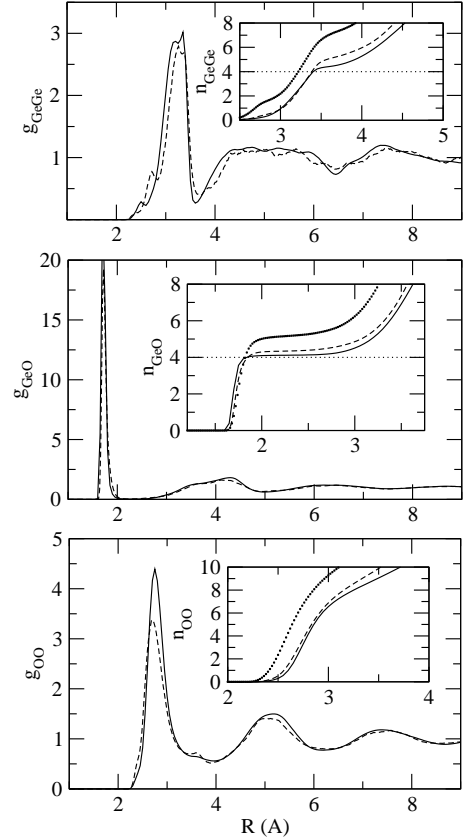}
\end{center}
\caption{Calculated partial pair correlation functions 
in vitreous germania at 300K at ordinary density $\rho_g=3.66~g.cm^{-3}$ 
(solid line) and in a permanently densified
system ($\rho=4.5~g.cm^{-3}$, dashed
line). The insert represent the corresponding running coordination
number $n_{ij}(R)$ together with results from a system under
$16.6~GPa$ pressure (dots).}
\end{figure}
One has however to note that these measurements have been refined
but without any complete resolution of all the partials\cite{Price2003}. 
The principal features in the structure factors of $GeO_2$
are the peaks occuring at $Q_P=1.5-2\AA^{-1}$ (the first sharp diffraction
peak, FSDP, corresponding to a correlation length $L_1=2\pi/Q_P\simeq4.1\AA$). 
Peaks at $2.6\AA^{-1}$ and $4\AA^{-1}$ appearing in the
partial structure factors of the Ge-O and O-O can be associated with
chemical short range order (CRSO) and topological
short-range order as derived from scaling considerations\cite{r17}. For the
former, the similar height in $g_{GeGe}$ and $g_{OO}$ but with
opposite sign suggests the CRSO nature of the network. The
evolution of the total scattering function and the position of the
FSDP with respect to pressure is discussed below.
\par
The calculation of the pair correlation functions permits to extract
the simulated bond lengths which also agree with experimental
findings: $d_{Ge-Ge}=3.32\AA$, $d_{Ge-O}=1.72\AA$ and $d_{O-O}=2.81\AA$ (to be
respectively compared to the experimental values\cite{r11} of
$3.16\pm0.03\AA$, $1.73\pm0.03\AA$ and $2.83\pm0.05\AA$). Weak changes are found
when comparing the virgin and the permanently densified systems. The
most striking difference is found in the Ge-Ge correlator which
differs even at low distances whereas the loss of peak heights in
$g_{OO}$ produces a slight shift in the corresponding running
coordination number $n_{OO}$. The number of oxygen neighbours
around an germanium atom remains however unaffected by pressure
release. More dramatic is the change in structure in the pressurized
system at $16.6~GPa$ (dots in the inserts of Fig. 2) which show
substantial differences in the bond distances ($d_{Ge-Ge}=3.25~\AA$,
$d_{Ge-O}=1.75~\AA$ and $d_{O-O}=2.56~\AA$) and an increase of the number
of oxygen neighbours in the vicinity of a Ge atom (about 5 in the
insert representing $n_{GeO}$, see also Fig. 3). 
Evidence of supplementary atoms in
the first shell surrounding a central $GeO_{4/2}$ unit is also
provided by the increase of $n_{GeGe}$.
\par
\begin{figure}
\begin{center}
\epsfig{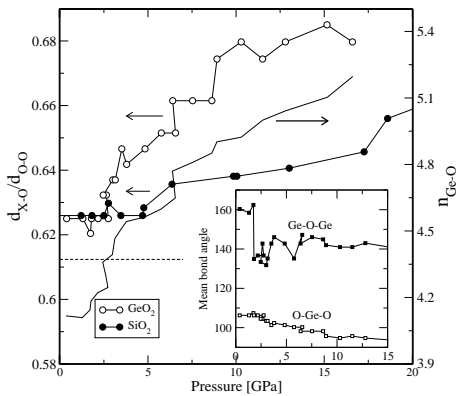}
\end{center}
\caption{Left axis: The distorion parameter of a regular $GeO/{4/2}$ as a
function of applied pressure(open circles). For comparison, the same
parameter for $SiO_2$ (filled circles). The dashed horizontal line
represents the value $\sqrt{3/8}$.The right axis represents the Ge-O
coordination number (solid line).}
\end{figure}
Using the present simulation, it is now possible to analyze in more
depth the low pressure behaviour. 
Focus is made on the distorsion of the tetrahedron parameter
defined by $\delta=d_{X-O}/d_{O-O}$ [X=Si,Ge] which provides a direct
measure of the effect of the pressure on the local structure of the
network. For an ideal tetrahedron, $\delta=\sqrt{3/8}$. The latter quantity
is of central interest when studying the flexibility of the glass
under pressure\cite{Dove} and the rigid unit modes in the context of
pressure induced rigidity\cite{Dove1,Dove2}. In Fig. 3 (left axis) are
represented the variation of the
tetrahedron parameter $\delta$ with pressure for both silica\cite{silica} and 
germania. At low pressure, $\delta$ remains almost constant suggesting that
the tetrahedral environment is preserved, slightly
higher however than the value of a perfect tetrahedron. For
$P\geq2.8~GPa$, there is increasing distorsion of
the tetrahedron in germania. The way of distorsion appears however to be
radically different as compared to the silica system. In the former it is
found a stepwise increase (most noticable from the jump of $\delta$ at
around $3~GPa$), in contrast with the more or less smooth increase of 
$\delta$ for the latter.
In the insert of Fig. 3 are shown
the variation of the mean bond angles O-Ge-O and Ge-O-Ge with respect
to pressure. For both the virgin and the permanently densified system,
the O-Ge-O bond angle peaks at around $109^o$. As pressure is
increased, the intratetrahedral bond angle decreases since the Ge-O
bond distances increases. The
constant value of the O-Ge-O bond angle at low pressure correlates of
course with the absence of distorsion of the basic tetrahedron. The
main feature provided by the angular analysis comes from the variation
of the
intertetrahedral mean bond angle Ge-O-Ge with pressure, which exhibits
a sharp drop
at around $1.8~GPa$, from $158^o$ to $135^o$ followed by a
stabilization at around $140^o$. Densification first applies on the
angles connecting the tetrahedra and preserves the latter. If pressure
keeps increasing, the distorsion of the tetrahedra sets in, which in
turn stabilizes more or less the Ge-O-Ge bond angle.
\par
\begin{figure}
\begin{center}
\epsfig{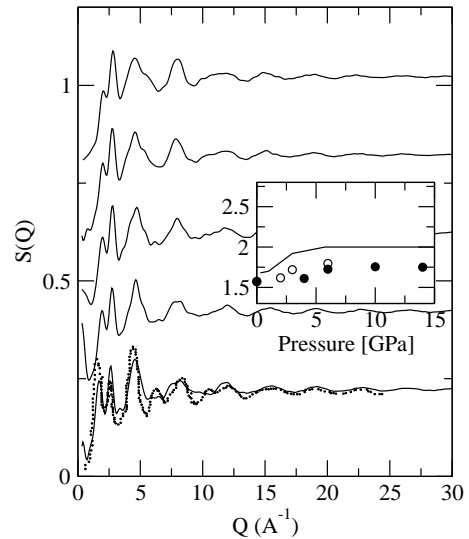}
\end{center}
\caption{Scattering factor $S(Q)$ for different applied pressures:
0,3.0, 5.76, 8.92 and 16.6GPa. The lower curve is the virgin system at
0GPa, compared to experiments [11]. The insert shows the simulated
position in $Q_P$ of the FSDP
with pressure (solid line). Experimental data are from [23](open circles) and
[9](filled circles).}
\end{figure}
Further interpretation is provided from the variation of the position
of the FSDP with respect to pressure. In Fig. 4 are represented the
scattering functions for different applied pressures which show a
global broadening of the peak at $4~\AA^{-1}$ whereas the position of
the FSDP at $1.5~\AA^{-1}$ is shifted to the higher values in $Q$,
already at very low pressure and even before the angular drop at
$P=1.8~GPa$. With increasing pressure, the calculated FSDP broadens
and becomes less intense, as currently observed from X-ray or neutron 
diffraction\cite{r11}.
It is worth to note
that the peak at $2.5~\AA^{-1}$ obtained in the present simulation is
only weakly observable in the experiments from neutron
diffraction\cite{Hannon} displayed in Fig. 4. However, the simulated
double peak distribution between $1.5$ and $2.5~\AA^{-1}$ has been observed by
different authors\cite{Waseda} (see also discussion in
\cite{r12}).  The evolution with pressure of the position $Q_P$ of the
FSDP is represented in the insert of Fig. 4. Both
experiments and simulation show that $Q_P$ already increases at low
pressures and then stabilizes at around $Q=1.7~\AA^{-1}$ ($\simeq2~\AA^{-1}$ in
the simulation). This suggests that intermediate range order is
immediately affected by the densification and then remain unaffected
with further densification.
\par
In conclusion, we have shown that simulated $GeO_2$ under pressure
shows several main features with applied pressure and pressure
release: i) a global increase
of the number of oxygen neighbours in the vicinity of a germanium
atom, ii) a stepwise change in the local
structure with applied pressure, made of a reduction of long range 
correlation (seen from the position of the FSDP), a sharp reduction of the
intertetrahedral bond angle and then a progressive distorsion of the
$GeO_{4/2}$ tetrahedron and iii) no noticable change in local
structure between a virgin and a permanently densified system. This
clearly draws the following picture: Pressure
applies on different length scales. With increasing magnitude,
densification is realized by a successive deformation of long range
structure, intermediate (angular) and finally short range
structure (tetrahedral).
\par
It is a pleasure to acknowledge instructive discussions with Yves
Guissani and Bertrand Guillot. The Laboratoire de Physique 
Th{\'e}orique des Liquides is Unit{\'e} Mixte de
Recherche n. 7600 du CNRS.

\ecols

\begin{thebibliography}{100}
\bibitem{detail} The pressure at $15.16~GPa$ has been obtained from a
densified system at $\rho=5.9~g.cm^{-3}$.
\end{thebibliography}
\end{document}